\documentclass[prd,twocolumn,showpacs,nofootinbib,preprintnumbers,superscriptaddress]{revtex4}
\usepackage{amsmath}
\usepackage{amsfonts}
\usepackage{graphicx}
\usepackage{hyperref}
\usepackage{mathbbol}

\def\be{\begin{equation}}
\def\ee{\end{equation}}
\def\bea{\begin{eqnarray}}
\def\eea{\end{eqnarray}}
\def\f{\frac}
\def\s{\sqrt}
\def\l{\left}
\def\r{\right}
\def\e{\epsilon}

\def\k{\kappa}
\def\a{\alpha}     
\def\b{\beta}
\def\no{\nonumber}

\begin{document}
\title{Slow-roll, acceleration, the Big Rip and WKB approximation in NLS-type formulation of scalar field cosmology}
\date{\today}

\vspace{2cm}

\author{Burin Gumjudpai}
  \affiliation{Centre for
Theoretical Cosmology, Department of Applied Mathematics and Theoretical Physics, University of Cambridge\\ Centre for Mathematical Sciences, 
Wilberforce Road, Cambridge CB3 0WA, United Kingdom}\email{B.Gumjudpai@damtp.cam.ac.uk}
 \affiliation{Fundamental Physics \&
Cosmology Research Unit, The Tah Poe Academia Institute (TPTP)\\
Department of Physics, Naresuan University, Phitsanulok 65000,
Siam}\email{buring@nu.ac.th}
\date{\today}

\begin{abstract}
Aspects of non-linear Schr\"{o}dinger-type (NLS) formulation of scalar (phantom) field cosmology on slow-roll, acceleration, WKB approximation 
and Big Rip singularity are presented. Slow-roll parameters for the curvature and barotropic density terms are introduced. We reexpress all 
slow-roll parameters, slow-roll conditions and acceleration condition in NLS form. WKB approximation in the NLS formulation is also discussed 
when simplifying to linear case. Most of the Schr\"{o}dinger potentials in NLS formulation are very slowly-varying, hence WKB approximation is 
valid in the ranges. In the NLS form of Big Rip singularity, two quantities are infinity in stead of three. We also found that approaching the 
Big Rip, $w_{\rm eff}\rightarrow -1 + {2}/{3q}$, $(q<0)$ which is the same as effective phantom equation of state in the flat case. 

\end{abstract}

\pacs{98.80.Cq}

\date{\today}

\vskip 1pc

\maketitle \vskip 1pc
\section{Introduction}
\label{sec:introduction} Cosmology with scalar field is one of today
research mainstreams. Although the scalar field has not yet been
observed, it is motivated from many ideas in high energy physics and
quantum gravities. Near future TeV scale experiments at LHC and
Tevatron might discover its existence. It has been widely accepted
in theoretical frameworks especially in model building of
contemporary cosmology that the field sources acceleration expansion
at early time, i.e. inflation, in order to solve horizon and
flatness problems \cite{inflation} and it also plays similar role in
explaining present acceleration observed and confirmed from cosmic
microwave background \cite{Masi:2002hp}, large scale structure
surveys \cite{Scranton:2003in} and supernovae type Ia
\cite{Riess:1998cb, Riess:2004nr,Astier:2005qq}. In the late
acceleration, it plays the role of dark energy (see Ref.
\cite{Padmanabhan:2004av} for reviews). Both inflation and
acceleration are convinced by recent combined results
\cite{Spergel:2006hy} with possibility that the scalar field could
be phantom, i.e. having equation of state coefficient $w_{\phi} <
-1$. The phantom equation of state is attained from negative kinetic
energy term in its Lagrangian density \cite{Caldwell:1999ew,
Melchiorri:2002ux}. Using BBN constraint of limit of expansion rate
\cite{Steigman:2007xt,Wright:2007vr} with most recent WMAP five-year
result \cite{Hinshaw:2008kr}, $w_{\phi,0} = -1.09 \pm 0.12$ at 68\%
CL.  While WMAP five-year result combined with Baryon Acoustic
Oscillation of large scale structure survey (from SDSS and 2dFGRS)
\cite{Percival:2007yw} and type Ia supernovae data (from HST
\cite{Riess:2004nr}, SNLS \cite{Astier:2005qq} and ESSENCE
\cite{WoodVasey:2007jb}) assuming dynamical $w$ with flat universe
yields $-1.38 < w_{\phi,0} < -0.86$ at 95\% CL and $w_{\phi,0} =
-1.12 \pm 0.13$ at 68\% CL. Although the phantom field has its room
from observation, in flat universe the idea suffers from unwanted
Big Rip singularity \cite{starobinsky:1999, Caldwell:2003vq}.
However there have been many attempts to resolve the singularity
from both phenomenological and fundamental inspirations
\cite{Sami:2005zc}.

Inflationary models in presence of other field behaving
barotropic-like apart from having only single scalar field were
considered such as in \cite{Chaicherdsakul:2006ui} where the scale
invariant spectrum in the cosmic microwave background was claimed to
be generated not only from fluctuation of scalar field alone but
rather from both scalar field and interaction between gravity to
other gauge fields such as Dirac and gauge vector fields. This is
similar to the situation in the late universe in which the
acceleration happens in presence of both dark matter fluid and
scalar fluid (as dark energy). Proposal of mathematical alternatives
to the standard Friedmann canonical scalar field cosmology with
barotropic perfect fluid, was raised, such as non-linear
Ermakov-Pinney equation \cite{Hawkins:2001zx, Williams:2005bp}.
There are also other applications of Ermakov-Pinney equation, for
example in \cite{Lidsey:2003ze}, a link from standard cosmology with
$k>0$ in Ermakov system to Bose-Einstein condensates was shown.
Another example is a connection from generalized Ermakov-Pinney
equation with perturbative scheme to generalized WKB method of
comparison equation \cite{Kamenshchik2006}. Recently a link from
standard canonical scalar field cosmology in
Friedmann-Lema\^{i}tre-Robertson-Walker (FLRW) background with
barotropic fluid to quantum mechanics is established. It was
realized from the fact that solutions of generalized Ermakov-Pinney
equation are correspondent to solutions of the non-linear
Schr\"{o}dinger-type equation, hereafter NLS equation
\cite{Williams:2005bp, D'Ambroise:2006kg}. Connection from the
NLS-type formulation to Friedmann scalar field cosmology formulation
is concluded in Ref. \cite{Gumjudpai:2007qq} where standard
cosmological quantities are reinterpreted in the language of quantum
mechanics assuming power-law expansion, $a \sim t^q$ and the phantom
field case is included. The quantities in the new form satisfies a
non-linear Schr\"{o}dinger-type equation. In most circumstance, the
scalar field exact solution $\phi(t)$ can be solved analytically
only when assuming flat geometry ($k=0$) and scalar field fluid
domination. When $k\neq 0$ with more than one fluid component, the
system is not always possible to solve analytically in standard
Friedmann formulation. Transforming standard Friedmann cosmological
quantities into NLS forms could help solving for the solution
\cite{Gumjudpai:2007bx,Phetnora:2008}.  In the NLS formulation, the
independent variable $t$ in standard formulation is re-scaled to
variable $x$. However, pre-knowledge of scale factor as function of
time, $a(t)$, must be presumed in order to express NLS quantities.
It is interesting to see the other features of field velocity,
$\dot{\phi}$, e.g. acceleration condition, slow-roll approximation,
written in NLS formulation. Mathematical tools such as WKB
approximation in quantum mechanics might also be interesting for
application in standard scalar field cosmology. It is worthwhile to
investigate this possibility. It is worth noting that
Schr\"{o}dinger-type equation in scalar field cosmology was
previously considered in different procedure to study inflation and
phantom field problems \cite{Chervon:1999}.

We introduce the NLS formulation in Sec. \ref{sec:NS}. The slow-roll conditions in both formulations are discussed in Sec. \ref{sec:slowroll} where
we define slow-roll parameters for barotropic fluid and curvature terms.
Then in Sec. \ref{sec:acc} we show acceleration conditions in NLS form. The WKB approximation is performed in Sec. \ref{sec:WKB}.
The NLS form of Big Rip singularity is in Sec. \ref{sec:BigRip} and finally conclusion is made in Sec. \ref{sec:con}.

\section{Scalar field cosmology in NLS formulation} \label{sec:NS}
Two perfect fluids are considered in FLRW universe: barotropic fluid
and scalar field. The barotropic equation of state is $p_{\gamma}
=w_{\gamma}\rho_{\gamma}$ with $w_{\gamma}$ expressed as $n$ where $
n = 3(1+w_{\gamma})$. The scalar field pressure obeys $p_{\phi} =
w_{\phi}\rho_{\phi}$. Total density and pressure of the mixture are
sum of the two components. Evolution of barotropic density is
governed by conservation equation$, \dot{\rho}_{\gamma} = - n H
\rho_{\gamma} $ with solution,
$\rho_{\gamma} = {D}/{a^{n}}\,,$
 where $a$ is scale factor, the dot denotes time derivative, $D\geq 0$ is a proportional
constant. Using scalar field Lagrangian density, $ \mathcal{L} =
(1/2)\epsilon \dot{\phi}^2 - V(\phi)$, i.e. minimally coupling to
gravity,
\bea \rho_{\phi} = \frac{1}{2} \epsilon \dot{\phi}^2 + V(\phi)\,,
\;\;\;\;\;\;\;\;
 p_{\phi}= \frac{1}{2} \epsilon \dot{\phi}^2 - V(\phi)\,.
\label{phanp}\eea The branch $\epsilon=1$ is for non-phantom case
and $\epsilon=-1$ is for phantom case \cite{Caldwell:2003vq}. Note
that the phantom behavior ($\rho_{\phi} < -p_{\phi}$) can also be
obtained in non-minimal coupling to gravity case
\cite{Boisseau:2000}. Dynamics of the field is controlled by
conservation equation \be \epsilon\left(\ddot{\phi} + 3H\dot{\phi}
\right) = -\frac{{\rm d}V}{{\rm d}\phi}\,. \label{phanflu} \ee
The spatial expansion of the universe sources friction to dynamics
of the field in Eq. (\ref{phanflu})
via the Hubble parameter $H$. The Hubble parameter is governed by Friedmann equation, %
\bea
H^2 &=& \frac{\kappa^2}{3}\rho_{\rm tot} - \frac{k}{a^2}\,,
\label{fr}
\eea
where here $\rho_{\rm tot} =
(1/2)\epsilon{\dot{\phi}}^2 + V + {D}/{a^{n}}$ and  by acceleration
equation,
 \bea
\frac{\ddot{a}}{a} &=&  -\frac{\kappa^2}{6} (\rho_{\rm tot}
 + 3p_{\rm tot})\,,  \label{ac}\eea 
which does not depend on $k$. This gives acceleration condition
  \bea
  p_{\rm tot} < -\frac{\rho_{\rm tot}}{3} \,.\label{acptot}
  \eea
Here $p_{\rm tot}= w_{\rm eff}\rho_{\rm tot}, \kappa^2 \equiv 8\pi G
= 1/M_{\rm P}^2$, $G$ is Newton's gravitational constant, $M_{\rm
P}$ is reduced Planck mass, $k$ is
spatial curvature and %
$ w_{\rm eff} = ({\rho_{\phi}w_{\phi} + \rho_{\gamma}
w_{\gamma}})/{\rho_{\rm tot}}$.
 Using these facts, it is straightforward to show that%
\bea   \epsilon  \dot{\phi}(t)^2 & = & -\frac{2}{\kappa^2} \left[ \dot{H} - \frac{k}{a^2}  \right] - \frac{n D}{3  a^n} \,, \label{phigr} \\
V(\phi) &=& \frac{3}{\kappa^2} \left[H^2 + \frac{\dot{H}}{3} +
\frac{2k}{3 a^2} \right] + \left(\frac{n-6}{6}\right)
\frac{D}{a^n}\,. \label{Vgr} \eea %
The Friedmann formulation of scalar field cosmology above can be
transformed to the NLS formulation as one defines NLS quantities
\cite{D'Ambroise:2006kg}, \bea
 u(x) &\equiv& a(t)^{-n/2}\,, \label{utoa} \\
 E &\equiv&  -\frac{\kappa^2 n^2}{12} D \,, \label{E} \\
P(x) &\equiv& \frac{\kappa^2 n}{4}a(t)^{n} \epsilon
\dot{\phi}(t)^2 \,. \label{schropotential} \eea
In the NLS formulation, there is no such analogous equations to
Friedmann equation or fluid equation since both of them are written
together in form of a non-linear Schr\"{o}dinger-type
equation\footnote{ NLS equation considerd here is only $x$-dependent
hence it is not partial differential equation with localized
soliton-like solution as in \cite{solitons}},
 \bea u''(x) + \left[E-P(x)\right]
u(x)
 = -\frac{nk}{2}u(x)^{(4-n)/n}\,,   \label{schroeq} \eea
where $'$ denotes ${\rm d}/{\rm d}x$. Independent variable $t$ is
scaled to NLS independent variable $x$ as $ x = \sigma(t), $ such
that
\bea \dot{x}(t)&=& u(x)\,, \label{dsigtou} \\
\phi(t) &=& \psi(x)\,.  \label{phitopsi}\eea
Using Eq. (\ref{schropotential}) and $\epsilon \dot{\phi}(t)^2 =
\epsilon \dot{x}^2\, \psi'(x)^2$, we get \cite{Gumjudpai:2007qq}
 \bea \epsilon\,\psi'(x)^2 = \frac{4}{ \kappa^2 n}\,P(x)\,, \eea
hence \bea \psi(x) =
\pm\frac{2}{\kappa\sqrt{n}}\int{\sqrt{\frac{P(x)}{\epsilon}}}\,{\rm
d}x\,  \,.
\label{phitoPx} %
\eea
Inverse function $\psi^{-1}(x)$ exists for $P(x) \neq 0$ and $n \neq
0$. In this circumstance, $ x(t) = \psi^{-1}\circ \phi(t)$ and the
scalar field potential, $V\circ \sigma^{-1}(x)$ and $\epsilon
\dot{\phi}(t)^2$ can be expressed in NLS formulation as
\bea \epsilon \dot{\phi}(x)^2 & =&  \frac{4}{\kappa^2 n}u u'' + \frac{2
k}{\kappa^2} u^{4/n} + \frac{4E}{ \k^2 n}u^2 = \f{4 P}{\k^2 n}u^2\,, \label{phidotu} \\
    V(x) &=&
\frac{12}{\kappa^2 n^2}( u')^2 - \frac{2 P}{\kappa^2 n}u^2  +
\frac{12 E }{\kappa^2 n^2}u^2 + \frac{3 k }{\kappa^2}u^{4/n}\,.
\label{vt} \eea
From Eqs. (\ref{phidotu}) and (\ref{vt}), we can find
\bea
\rho_{\phi} &=& \frac{12}{\kappa^2 n^2}( u')^2   +
\frac{12 E }{\kappa^2 n^2}u^2 + \frac{3 k }{\kappa^2}u^{4/n} \,,
\label{NLSrho_phi} \\
p_{\phi} &=& -\frac{12}{\kappa^2 n^2}( u')^2   +  \frac{4 P}{\kappa^2 n}u^2
-\frac{12 E }{\kappa^2 n^2}u^2 - \frac{3 k }{\kappa^2}u^{4/n} \,.
\label{NLSrho_p}
\eea
We know that $\rho_{\gamma} = Du^2 = -12Eu^2/(\k^2 n^2)$ from Eq. (\ref{E})
 and the barotropic pressure is $p_{\gamma} = [(n-3)/3]\rho_{\gamma}$, therefore
 \bea
   \rho_{\rm tot} & = & \frac{12}{\kappa^2 n^2}( u')^2   +
 \frac{3 k }{\kappa^2}u^{4/n} \,, \label{rhoNLS} \\
   p_{\rm tot} & = & -\frac{12}{\kappa^2 n^2}( u')^2   +\f{4 u^2}{\k^2 n}\l[P-E\r] -
 \frac{3 k }{\kappa^2}u^{4/n} \,. \eea
 Using the Schr\"{o}dinger-type equation (\ref{schroeq}), then
 \bea  p_{\rm tot} & = & -\frac{12}{\kappa^2 n^2}( u')^2   +\f{4 }{\k^2 n}u u'' -
 \frac{ k }{\kappa^2}u^{4/n}. \label{pNLS}
 \eea

\section{Slow-roll conditions} \label{sec:slowroll}
\subsection{Slow-roll conditions: flat geometry and scalar field domination}
In flat universe with scalar field domination ($k=0, \rho_{\gamma} =
0$), the Friedmann equation  $ H^2  = {\kappa^2}\rho_{\phi}/{3}\,, $
together with the Eq. (\ref{phanflu}) yield  $ \dot{H} =
-{\kappa^2}\dot{\phi}^2\epsilon/{2}\,$. For $\epsilon = -1$, we get
$\dot{H} > 0$ and  %
\be%
 0 < aH^2 < \ddot{a}\,, %
\ee %
i.e. the acceleration is greater than speed of expansion per Hubble
radius, $\dot{a}/cH^{-1}$. On the other hand, for $\epsilon =1$, we
get $\dot{H} < 0$ and
\be%
 0  < \ddot{a} < aH^2 \,. %
\ee %
Slow-roll condition in \cite{Liddle:1992wi,liddlebook} assumes
negligible kinetic term hence $|\epsilon\dot{\phi}^2/2| \ll
V(\phi)$, therefore $\rho_{\phi} \simeq V(\phi)$ hence $H^2 \simeq
\kappa^2V/3$. With this approximation, %
\bea %
H^2 = -\frac{\dot{H}}{3} + \frac{\kappa^2}{3}V\,, %
\;\;\Rightarrow\;\; H^2 \simeq -\frac{\dot{H}}{3} + H^2 \,.
\label{expresso} %
\eea %
 This results in an
approximation $ |\dot{H}| \ll H^2$ from which the slow-roll
parameter, \be \varepsilon \equiv -\f{\dot{H}}{H^2} \ee is defined.
Then the condition $|\epsilon\dot{\phi}^2/2| \ll V(\phi)$ is
equivalent to $|\varepsilon| \ll 1$, i.e.
 $-1 \ll \varepsilon < 0 $ for phantom field case and $0 <
\varepsilon \ll 1 $ for non-phantom field case.  For the non-phantom
field, this condition is necessary for inflation to happen (though
not sufficient) \cite{Liddle:1992wi} but for the phantom field case,
the slow-roll condition is not needed because the negative kinetic
term results in acceleration with $w_{\phi} \leq -1 $. The other
slow-roll parameter is defined by balancing magnitude of the field
friction and acceleration terms in Eq. (\ref{phanflu}). This is
independent of $k$ or $\rho_{\gamma}$. When friction dominates
$|\ddot{\phi}| \ll |3H\dot{\phi}|$, then \be \eta \equiv
-\f{\ddot{\phi}}{H\dot{\phi}}\ee is defined \cite{Liddle:1992wi}.
The condition is then $|\eta| \ll 1$ and the fluid equation is
approximated to %
$ \dot{\phi} \simeq - {V_{\phi}}/{3\epsilon H}\, %
$ %
 which allows the field to roll up the hill when $\e=-1$.
Using both conditions, e.g. $|\epsilon\dot{\phi}^2/2| \ll V$ and
$|\ddot{\phi}| \ll |3H\dot{\phi}|$ together, one can derive %
$ %
\varepsilon = ({1}/{2\kappa^2 \epsilon}) ({V_{\phi}}/{V})^2 $ and $
\eta = (1/\kappa^2) ({V_{\phi\phi}}/{V})\,$ as well-known.
%
\subsection{Slow-roll conditions: non-flat geometry and non-negligible barotropic density}
\subsubsection{Friedmann formulation}
When considering the case of  $k \neq 0$ and $\rho_{\gamma} \neq 0$, then %
\be %
\dot{H} = - \frac{\k^2}{2}\dot{\phi}^2 \e + \f{k}{a^2} - \f{n \k^2
}{6} \f{D}{a^n} \,. \label{coffeeMixdot} \ee We can then write slow-roll
condition as: $|\k^2 \epsilon\dot{\phi}^2/6| \ll (\k^2 V/3) -
(k/a^2) + (\k^2 D/3 a^n)$ and hence $H^2 \simeq (\k^2 V/3) + (\k^2
D/3 a^n) - (k/a^2)$. Using this approximation and Eq.
(\ref{coffeeMixdot}) in (\ref{fr}),  %
\be %
H^2 \simeq -\frac{\dot{H}}{3} + \f{k}{3a^2} - \f{n \k^2}{18}\f{D}{a^n} + H^2 \,, %
\ee which implies $  |-({\dot{H}}/{3}) + ({k}/{3a^2}) - ({n \k^2
D}/{18 a^n}) | \ll H^2 $. We can reexpress this slow-roll condition as %
\bea %
|\varepsilon + \varepsilon_k + \varepsilon_D| \; \ll \; 1\,, \label{SUMep} \eea where $\varepsilon_k \equiv k/a^2 H^2$ and $\varepsilon_D \equiv 
-n \k^2 D/6 a^n H^2 $. Another slow-roll parameter $\eta$ is defined as $\eta \equiv -\ddot{\phi}/H\dot{\phi}$, i.e. the same as the flat scalar 
field dominated case since the condition $|\ddot{\phi}| \ll |3H\dot{\phi}|$ is derived from fluid equation of the field which is independent of 
$k$ and $\rho_{\gamma}$.
\subsubsection{NLS formulation}
In NLS formulation, the Hubble parameter takes the form  %
\bea
H =-\f{2}{n} u' \,, \label{NLSH} \eea
with
\bea \dot{H} \, =\,  -\f{2}{n} u u''
             \, = \,  \f{2}{n} u^2 \l[E-P(x)\r] + k u^{4/n} \,.
\label{NLSHDOT}  \eea The slow-roll condition
$|\epsilon\dot{\phi}^2/2| \ll V$ using Eqs. (\ref{schropotential})
and (\ref{vt}) in NLS form, is then
\bea
|P(x)| \ll
\frac{3}{n}\left[ \l( \f{u'}{u}\r)^2 + E \right] +
\f{3}{4}k\, n\, u^{(4-2n)/n}\,. \label{NLSa} %
\eea If the absolute sign is not used, the condition is then
$\epsilon\dot{\phi}^2/2 \ll V $, allowing fast-roll negative kinetic
energy. Then Eq. (\ref{NLSa}), when combined with the
NLS equation (\ref{schroeq}), yields %
\bea u''\ll \f{3}{n}\f{u'\,^2}{u} + \l( \f{3}{n}-1 \r)E u +
\frac{kn}{4} u^{(4-n)/n}\,.\eea Friedmann formulation analog of this condition can be obtained simply by using
Eqs (\ref{phigr}) and (\ref{Vgr}) in the condition.
Consider another aspect of slow-roll in the fluid equation, the field
acceleration can be written in NLS form:
\be \ddot{\phi} =  \frac{2 P u u' + P' u^2}{\k \s{P \e n}}
\,,%
 \ee while the friction term in NLS form is
 \be
3H\dot{\phi} = - \frac{12 u' u }{n \k}  \s{\frac{P}{\e n }}\,.
 \ee
The second slow-roll condition, $|\ddot{\phi}| \ll |3H\dot{\phi}| $
hence corresponds to %
\be%
 \l| \f{P'}{P} \r|\: \ll \: \l|-2 \l(\f{6+n}{n}\r)  \frac{u'}{u} \r|\,. %
\ee This condition yields the approximation $3H \e \dot{\phi}^2
\simeq -{\rm d }V/{\rm d} \phi$.  Using Eqs. (\ref{phidotu}),
(\ref{vt}), (\ref{NLSH}) and (\ref{NLSHDOT}), one can express the approximation, $3H \e \dot{\phi}^2
\simeq -{\rm d }V/{\rm d} \phi$, in NLS form as%
\bea %
\frac{P'}{P} \: \simeq \:  - \frac{2u'}{u} \:= \: n H a^{n/2}\,.
\eea
and finally the slow-roll parameters $\varepsilon$, $\varepsilon_k$ and $\varepsilon_D$, introduced previously, become
\bea \varepsilon  =    \f{n u u''}{2 {u'}^2}
\,, \;\;\;\;
\varepsilon_k  =   \f{n^2 k u^{4/n}}{{4 u'}^2} \,, \;\;\;\;
\varepsilon_D   =  \f{n E}{2}\l(\f{u}{u'}\r)^2 \,,  \label{NLSslowroll}
 \eea
in NLS form. With help of NLS equation (\ref{schroeq}), summation of
the slow-roll parameters takes simple form, \bea \varepsilon_{\rm
tot} = \varepsilon + \varepsilon_k + \varepsilon_D =
\f{n}{2}\l(\f{u}{u'} \r)^2  P(x)\,. \eea Finally the slow-roll
condition, $|\varepsilon_{\rm tot}| \ll 1$ (Eq. (\ref{SUMep})), in
NLS form, is \bea \l| \l(\f{u}{u'} \r)^2  P(x) \r| \ll 1\,.
\label{NLSepcon} \eea Another slow-roll parameter  $\eta =
-\ddot{\phi}/H\dot{\phi}$ can be found as follow. First considering
$\psi(x) = \phi(t)$ (Eq. (\ref{phitopsi})), using relation ${\rm
d}/{\rm d}t = \dot{x}\,{\rm d}/{\rm d} x$ and Eq. (\ref{NLSH}), we
obtain \be \eta = \f{n}{2} \l(\f{u}{u'} \f{\psi''}{\psi'} +1  \r)
\,. \ee The Eq. (\ref{phitoPx}) yields \bea \psi'  =  \pm \f{2}{\k }
\s{\f{P}{n \e}} \;\;\;\;\;{\rm and }\;\;\;\;\; \psi'' =  \pm
\f{P'}{\k\s{n P \e}}\,. \eea Hence \be \eta = \f{n}{2} \l(\f{u}{u'}
\f{P'}{2P} +1  \r)\,. \label{NLSeta} \ee At last, the slow-roll
condition $|\eta| \ll 1$ then reads \be \l|  \f{u}{u'} \f{P'}{2P} +1
\r| \, \ll \, 1\,. \label{NLSetacon} \ee

\section{Acceleration condition} \label{sec:acc}
The slow-roll condition is useful for non-phantom field because it is a necessary condition for inflating acceleration.
However, in case of phantom field, the kinetic term is always negative and could take any large negative values hence
slow-roll condition is not necessary for acceleration condition. More generally, to ensure acceleration, the Eq. (\ref{ac})
must be positive.
It is straightforward to show that, obeying acceleration condition, $\ddot{a} > 0 $, the Eq. (\ref{acptot}),
takes the form, %
\bea %
\epsilon \dot{\phi}(x)^2 &<& - \left(\frac{n-2}{2}\right) \frac{D}{a^n} + V\,. \label{accongr} %
\eea
 With  Eqs. (\ref{utoa}), (\ref{E}),
(\ref{schropotential}) and (\ref{vt})), the acceleration condition
(\ref{accongr}) in NLS-type formulation is
\bea %
E-P &>& - \frac{2}{n} \left(\frac{u'}{u}\right)^2    - \frac{nk}{2}\left(\frac{u^{2/n}}{u}\right)^2 \,. \label{acconNLS} %
\eea With help of non-linear Schr\"{o}dinger-type equation (\ref{schroeq}), it is simplified to %
\bea u'' < \frac{2}{n} \frac{{u'}^2}{u}\,. \label{acconNLS2} \eea Using Eqs. (\ref{NLSH}) and (\ref{NLSHDOT}), the acceleration condition is just
 $\varepsilon < 1$ without using any slow-roll assumptions.

\section{WKB Approximation} \label{sec:WKB}
WKB approximation can be assumed when the coefficient of highest-order derivative term in the Schr\"{o}dinger equation is small
or when the potential is very slowly-varying.
The Eq. (\ref{schroeq}), when $k=0$, is linear. It is then
\bea
-\f{1}{n} u'' + \l[\tilde{P}(x) -\tilde{E}\r] u = 0 \,. \label{WKB1}
\eea
where $\tilde{P}(x)\equiv P(x)/n$ and $\tilde{E}\equiv E/n $.
For a slowly-varying $P(x)$ with assumption of $n \gg 1$,  the solution of Eq. (\ref{WKB1}) can be written as $u(x) \simeq  A \exp[\pm i n W_0(x)]$,
where $W_0(x) = W(x_0)$ is the lowest-order term in Taylor expansion of the function $W(x)$ in $(1/n)$ about $x=x_0$,
\bea
W(x) = W(x_0) + W'(x_0) \f{(x-x_0)}{n} + \ldots \,. \label{W}
\eea
 Then an approximation
\bea W(x) = \pm \f{1}{n}\int_{x_1}^{x_2} k(x)\,{\rm d}x \simeq
W_0(x) \,, \label{WKB} \eea is made in analogous to the method in
time-independent quantum mechanics. The Schr\"{o}dinger wave number
is hence \bea k(x) = \f{2\pi}{\lambda(x)} = \sqrt{{n}\l[ \tilde{E} -
\tilde{P}(x)\r]}\,,  \label{wavenum} \eea and small variation in
$\lambda(x)$ is \bea \f{\delta\lambda}{\lambda(x)} = \l| \f{\pi
\tilde{P}'}{\s{n} \l[{\tilde{E} - \tilde{P}(x)}\r]^{3/2} }  \r| =
\l| \f{\pi {P}'}{ \l[{{E} - {P}(x)}\r]^{3/2} }  \r|   \,. \eea For
WKB approximation, ${\delta\lambda}/{\lambda(x)}\ll 1$. In real
universe, we have $n=3$ (dust) or $n=4$ (radiation) which is not
much greater than one. However, if considering a range of very
slowly-varying potential, $P' \simeq 0$ implying $\delta P/\delta x
\sim 0$, hence $\delta k/\delta x \sim 0 \sim  W'(x) $. Therefore
$W(x) \simeq W_0(x)$ still holds in this range. Since $u(x) =
a^{-n/2}$, using WKB approximation, we get \bea a \sim A \exp\l[{\pm
(2/n) i \int_{x_1}^{x_2} \sqrt{E-P(x)} } \,{\rm d} x \r]\,,
\label{NLSWKBa} \eea where $A$ is a constant. Examples of
Schr\"{o}dinger potentials for exponential, power-law and phantom
expansions are derived in
\cite{Gumjudpai:2007bx,Gumjudpai:2007qq,Phetnora:2008}. These
potentials are steep only in some small particular region but very
slowly-varying in most regions, especially at large value of $|x|$
which are WKB-well valid.

\section{Big Rip singularity} \label{sec:BigRip}
When the field becomes phantom, i.e. $\e = -1$, in a flat FRLW universe it leads to future Big Rip singularity\cite{starobinsky:1999, 
Caldwell:2003vq}. In flat universe, when $w_{\rm eff} < -1$, i.e. being phantom, the expansion obeys $a(t) \sim (t_{\rm a} - t)^q$, where $q 
= 2/3(1+w_{\rm eff}) < 0$ is a constant in time and $t_{\rm a}$ is a finite time\footnote {The relation $q = 2/3(1+w_{\rm eff}) < 0$ holds only 
when $k=0$.}. The NLS phantom expansion was studied in Ref. \cite{Phetnora:2008} with inclusion of non-zero $k$ case. Therein, the same 
expansion function is assumed with constant
 $q<0$ and $x$
is related to cosmic time scale, $t$ as %
$ x(t) = ({1}/{\beta})\,(t_{\rm a} - t)^{-\beta} + x_0\,, $ so that 
 $
 u(x)  = \left[ \beta (x-x_0) \right]^{\alpha}\,.  
 $ 
 Here $\alpha \equiv {qn/(qn-2)}$ and $\beta \equiv {(qn-2)/2}$ with conditions $0 < \alpha <1$ and $\beta < -1$ since $n > 0$ always. 
 The first and second $x$-derivative of $u$ are\footnote{Note that 
 $(x-x_0)$ and $\beta$ are negative hence $(x-x_0)^{\a} $, $\beta^{\a}$, $(x-x_0)^{\a-1} $ and $\beta^{\a-1}$ are imaginary.} 
 \bea
 u'(x) &=& \a \beta [\beta(x-x_0)]^{\a-1}\,, 
 \\ u''(x) & = &   \a (\a -1) \beta^2 [\beta(x-x_0)]^{\a-2}  \,,
 \eea
where exponents $\a-1$ and $\a-2$ are always negative. Using Eqs. (\ref{rhoNLS}) and (\ref{pNLS}), then 
\bea 
\rho_{\rm tot} &=& \f{12 \a^2 \b^2 }{\k^2 n^2} [ \b(x-x_0)  ]^{2(\a-1)} + \f{3k}{\k^2}[ \b(x-x_0)  ]^{4\a/n}\,, \nonumber \\   \\
p_{\rm tot} &=&  \f{4  \b^2 }{\k^2 n} [ \b(x-x_0)  ]^{2(\a-1)} \l[ \l(1-\f{3}{n} \r)\a^2 -\a \r] \nonumber \\ & &- \f{k}{\k^2}[ \b(x-x_0)  ]^{4\a/n}\,.  \\
            &=& \f{4 {u'}^2}{\k^2 n} \l[ \l(1-\f{3}{n} \r) - \f{1}{\a} \r]   - \f{k}{\k^2}[ \b(x-x_0)  ]^{4\a/n} \,. 
\eea
The Big Rip: $(a, \rho_{\rm tot}, |p_{\rm tot}|) \rightarrow \infty$ happens when $t\rightarrow t_{\rm a}^{-}$. In NLS formulation, if $a 
\rightarrow \infty$, then $u\rightarrow 0^{+}$ (Eq. (\ref{utoa})).  From above, we see that conditions of the Big Rip singularity are 
\bea
t\rightarrow t_{\rm a}^{-}\:\;\; &\Leftrightarrow&\;\;\;\;\;\;\;\;\;  x \rightarrow x_0^{-}\,, \no \\
a \rightarrow \infty \;\;\; &\Leftrightarrow&\;\;\;\; u(x) \rightarrow 0^{+} \,, \no \\
\rho_{\rm tot} \rightarrow \infty\;\;\; &\Leftrightarrow&\;\;\;  u'(x) \rightarrow \infty \,, \no \\
|p_{\rm tot}| \rightarrow \infty\;\;\; &\Leftrightarrow&\;\;\; u'(x)
\rightarrow \infty\,. \label{NLSBigRip} \eea 
The effective
equation of state $w_{\rm eff} = p_{\rm tot}/\rho_{\rm tot}$ can
also be stated in NLS language as a function of $x$. Approaching the Big
Rip, $x \rightarrow x_0^{-}$ and the effective equation of state approaches a value 
\bea
\lim_{x\rightarrow x_0^{-}} w_{\rm eff} =  \f{n}{3}\l(1-\f{1}{\a} \r)-1  = -1 + \f{2}{3q}\,,\eea
which is similar to the equation of state in flat case. 
\section{Conclusions} \label{sec:con}
We feature cosmological aspects of NLS formulation of scalar field
cosmology such as slow-roll conditions, acceleration condition and
the Big Rip. We conclude these aspects in standard Friedmann
formulation before deriving them in the NLS formulation. We consider
a non-flat FRLW universe filled with scalar (phantom) field and
barotropic fluid because, in presence of barotropic fluid density,
the NLS-type formulation is consistent \cite{Gumjudpai:2007bx}. We
obtain all NLS version of slow-roll parameters, slow-roll conditions
and acceleration condition. This provides such analytical tools in
the NLS formulation. For phantom field, due to its negative kinetic
term, the slow-roll condition is not needed. When the NLS system is
simplified to linear equation (this happens when $k=0$.) as in
time-independent quantum mechanics, we can apply WKB approximation
to the problem.  When $n\gg 1$, the wave function is semi-classical
which is suitable for the WKB approximation. However, this does not
work since physically $n$ can not be much greater than unity, i.e.
$n=3$ for dust and $n=4$ for radiation. However, the WKB
approximation can still be well-valid in a range of very
slowly-varying Schr\"{o}dinger potentials $P(x)$ which were
illustrated in
\cite{Gumjudpai:2007qq,Gumjudpai:2007bx,Phetnora:2008}. Using the  
WKB approximation, we obtain approximated scale factor function (Eq.
(\ref{NLSWKBa})). In a flat universe with phantom expansion, the Big
Rip singularity is its final fate.  When the Big Rip happens, three
quantities ($a(t), p(t) \rm and \;\, \rho(t)$) become infinity.
Rewriting the singularity in NLS form (Eq. (\ref{NLSBigRip})), we
can remove one infinite (see Eq. \ref{NLSBigRip}). We found that at
near the Big Rip, $w_{\rm eff}\rightarrow -1 + {2}/{3q}$ where $q<0$ is a constant exponent of the expansion $a(t) \sim (t_{\rm a} - t)^q$. 
This limit is the same as the effective phantom equation of state in the case $k=0$.

\section*{Acknowledgments} The author is supported
as a TRF Research Scholar by the Thailand Research Fund. This work is also supported by Naresuan University's Overseas Visiting Postdoctoral 
Research Fellowship and the Centre for Theoretical Cosmology, D.A.M.T.P. He thanks Anne-Christine Davis, Stephen Hawking and Neil Turok for 
their support.

\end{document}